\documentclass[cameraready]{Interspeech}
\usepackage{pifont}
\usepackage{amsmath,amssymb}
\usepackage{multirow}
\newcommand{\cmark}{\ding{51}}
\newcommand{\xmark}{\ding{55}}
\newcommand{\tightline}{\par\vspace{-5pt}\noindent\rule{\linewidth}{0.4pt}\vspace{-2pt}\par}
\title{Disentangling Reasoning in Large Audio-Language Models for Ambiguous Emotion Prediction}

\author[affiliation={1}, equalcontribution] {Xiaofeng}{Yu}
\author[affiliation={2}, equalcontribution] {Jiaheng}{Dong}
\author[affiliation={2}] {Jean}{Honorio}
\author[affiliation={3}] {Abhirup}{Ghosh}
\author[affiliation={1}] {Hong}{Jia}
\author[affiliation={2}] {Ting}{Dang}

\address{
    $^1$ University of Auckland, New Zealand \\
    $^2$ The University of Melbourne, Australia \\
    $^3$ University of Birmingham, United Kingdom
}

\email{xyu724@aucklanduni.ac.nz, jiahengd@student.unimelb.edu.au}


\usepackage{comment}
\usepackage{soul, xcolor}

\begin{document}

\maketitle

\begin{abstract}
Speech emotion recognition plays an important role in various applications. However, most existing approaches predict a single emotion label, oversimplifying the inherently ambiguous nature of human emotional expression. 
Recent large audio–language models show promise in generating richer outputs, but their reasoning ability for ambiguous emotional understanding remains limited. In this work, we reformulate ambiguous emotion recognition as a distributional reasoning problem and present the first systematic study of ambiguity-aware reasoning in LALMs. Our framework comprises two complementary components: an ambiguity-aware objective that aligns predictions with human perceptual distributions, and a structured ambiguity-aware chain-of-thought supervision that guides reasoning over emotional cues. Experiments on IEMOCAP and CREMA-D demonstrate consistent improvements across SFT, DPO, and GRPO training strategies.
\end{abstract}

\section{Introduction} 
Speech emotion recognition (SER) plays an important role in human–computer interaction~\cite{cai2020multi, wadley2022future, singla2024emotion}, conversational agents~\cite{zadeh2023adaptive}, and mental health applications~\cite{sharma2023real}. Despite recent advances, most SER systems are trained to predict a single discrete emotion category, which oversimplifies the complexity of human emotional expression. In practice, emotions are often ambiguous and mixed rather than belonging to a single class. This discrepancy between computational modeling and human emotional perception motivates the need for more expressive and uncertainty-aware emotion understanding paradigms.

Recent large audio–language models (LALMs) have demonstrated promising capabilities in generating richer textual outputs beyond single-label classification~\cite{halim2025token, wenda}. Prior work shows that LALMs can capture multiple emotional cues from speech and generate distributional emotion predictions~\cite{halim2025token}. However, their reasoning ability under highly ambiguous emotional conditions remains limited. 
While humans naturally reason under emotion ambiguity by weighing multiple cues and forming probabilistic judgements~\cite{scherer2003vocal, schirmer2017emotion}, current LALMs struggle to emulate.

Enhancing reasoning in LALMs has attracted increasing attention, particularly through chain-of-thought (CoT) prompting~\cite{audio-cot} and reinforcement learning (RL)-based post-training~\cite{wen2025sari, diao2025soundmind}. For example, supervised fine-tuning (SFT) with CoT supervision~\cite{zhifei2025audio}, or preference optimization methods such as Direct Preference Optimization (DPO)~\cite{DPO} and RL methods of Group Relative Policy Optimization (GRPO)~\cite{GRPO} show great promises. However, these studies mainly focus on deterministic speech understanding tasks such as audio question answering (AudioQA) that selects a single correct answer. In contrast, ambiguous emotion reasoning is inherently distributional: multiple emotional interpretations may simultaneously be plausible, and can be represented as a soft label or probability distribution over emotion classes (e.g., 40\% happy, 60\% surprised). 
Therefore, the challenge is to improve reasoning quality and also avoid premature collapse to a single deterministic emotional interpretation when uncertainty is present.

To address this gap, we reformulate ambiguous emotion recognition as a distributional reasoning problem and present the first systematic study of ambiguity-aware reasoning in LALMs. An effective model must (i) preserve affective uncertainty at the decision level and (ii) perform structured reasoning over emotional ambiguity, coherently integrating subtle and heterogeneous emotional evidence. Our framework proposes two key components to tackle these: (i) an ambiguity-aware objective that aligns predicted emotion distributions with human perceptual distributions via forward KL regularization to prevent affective collapse; and (ii) structured ambiguity-aware CoT supervision to guide evidence of emotion ambiguity integration before prediction. The framework is plug-and-play and compatible with different post-training strategies, evaluated across SFT, DPO, and GRPO. 
Our contributions are summarized:
\begin{itemize}
    \item We propose the first systematic study of ambiguity-aware reasoning in LALMs
    
\item We design two complementary objectives to support ambiguity-aware learning: an ambiguity-aware objective and structured ambiguity-aware CoT supervision.
    \item We evaluate the proposed paradigm across multiple post-training strategies on IEMOCAP and CREMA-D dataset, demonstrating effectiveness of both objectives.
\end{itemize}

By disentangling decision-level uncertainty modeling from reasoning enhancement, our work provides new insights into ambiguity-aware emotion understanding in LALMs.
\section{Related Work} 

\noindent\textbf{Ambiguous Speech Emotion Recognition. } Early studies on ambiguous emotion recognition include modelling emotions ambiguity using soft labels instead of categorical hard labels~\cite{Interpreting_ambiguous_emotional_expressions}, capturing ambiguity by simulating multiple annotators with multiple classifiers~\cite{Multi-classifier}, or handling ambiguous emotions as out-of-distribution cases~\cite{wu2024handling}. With the emergence of LALMs, several studies investigates ambiguity-aware emotion recognition. For instance,~\cite{halim2025token} examines whether LALMs implicitly encode emotion ambiguity by analyzing token-level prediction distributions. Another study~\cite{wenda} proposes augmenting multi-annotator labels with synthetic annotations generated by LALMs. 
However, these approaches 
ignores improving the reasoning processes of LALMs for ambiguous emotion perception.

\noindent\textbf{Reasoning Capabilities of LALMs. }
 Recent work on improving reasoning in LALMs mainly falls into two categories: chain-of-thought
 (CoT) and reinforcement learning (RL) based approaches.
CoT-based methods aim to teach step-by-step reasoning. For example, Audio-CoT~\cite{audio-cot} explores prompt-based CoT elicitation, while Audio-Reasoner~\cite{zhifei2025audio} further trains models to follow structured reasoning routines before producing answers. However, these methods typically employ CoT patterns designed for deterministic tasks (e.g., AudioQA), where the goal is to produce a single correct answer, and do not explicitly model emotional ambiguity.

RL-based approaches instead optimize reasoning behaviors through reward-driven learning. SARI~\cite{wen2025sari} and SoundMind~\cite{diao2025soundmind} improve reasoning capability using GRPO and REINFORCE++~\cite{hu2025reinforce++} respectively. However, these reasoning frameworks are still tailored to certain tasks where the objective is to identify a single correct answer. It remains unclear what post-training objectives and strategies are essential for ambiguous emotion reasoning.

\section{Methods}
\begin{figure*}[t]
  \centering
  \includegraphics[width=0.9\textwidth]{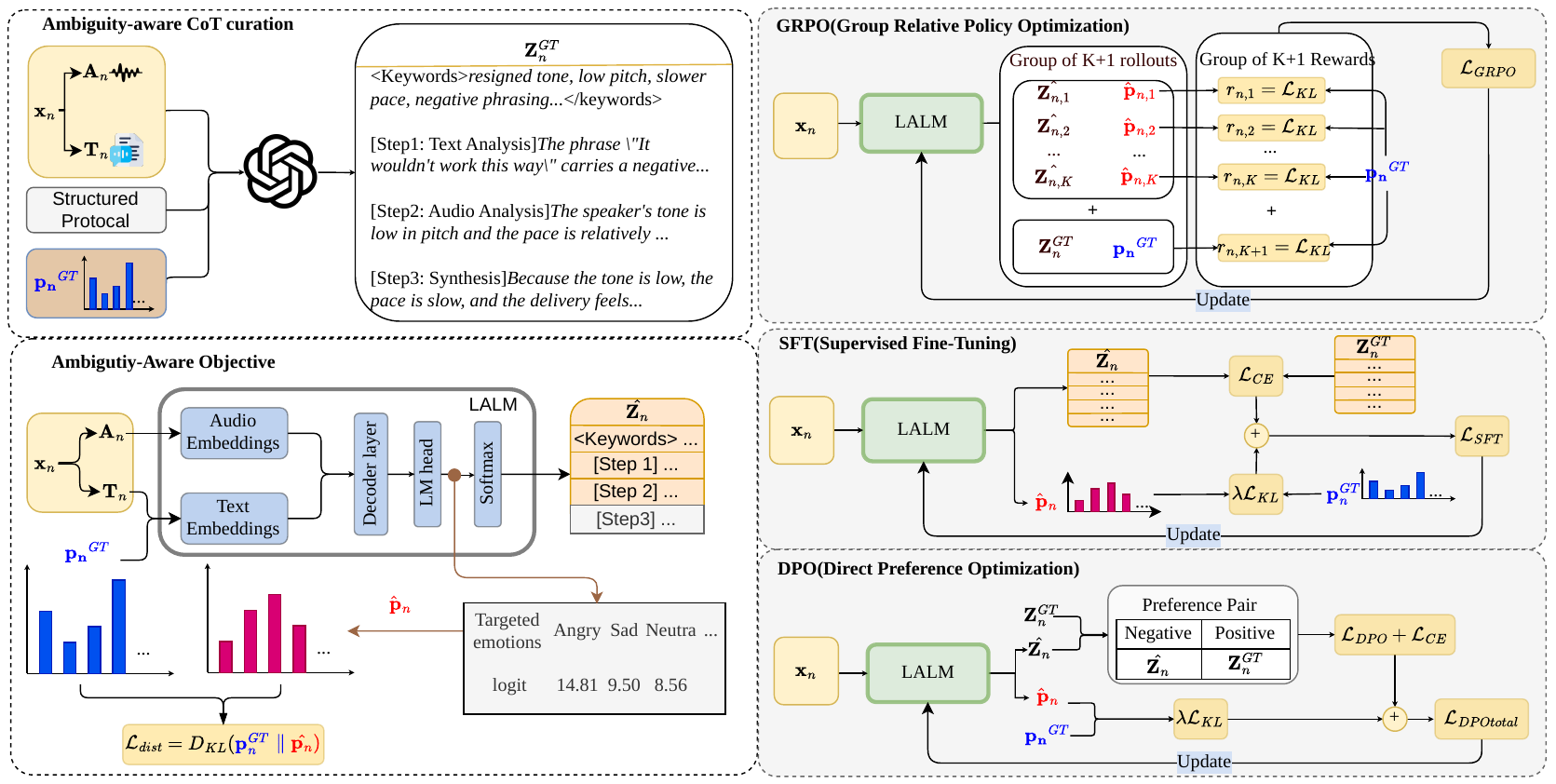}
  \vspace{-8pt}
  \caption{Overview of the proposed ambiguity-aware objectives and Plug-and-play Learning Paradigms.}
  \label{fig:workflow}
  \vspace{-10pt}
\end{figure*}

\vspace{-3pt}
\subsection{Problem Formulation}
We formulate emotion understanding as an ambiguity-aware reasoning task. For each speech utterance $n$, the input is a multimodal pair $x_n = \{A_n, T_n\}$, where $A_n$ denotes the acoustic signal and $T_n$ its transcript. To capture perceptual subjectivity, each sample is annotated with (i) a ground-truth emotion distribution $\mathbf{p}_n^{GT} \in \mathbb{R}^C$ over $C$ categories and (ii) a reasoning trajectory $\mathbf{Z}_n^{GT}$:
\begin{equation}
p^{GT}_{n,i} = \frac{m_{n,i}}{\sum_{j=1}^{C} m_{n,j}}, 
\quad 
\mathbf{Z}_n^{GT} = (z_n^{(1)}, \dots, z_n^{(L)})
\end{equation}
where $m_{n,i}$ denotes the number of annotators selecting class $i$, and $z_n^{(\ell)}$ represents an intermediate reasoning step. 
Unlike single-label classification, $\mathbf{p}_n^{GT}$ encodes collective emotion ambiguity as a soft distribution (e.g., 70\% happy and 30\% surprised). The trajectory $\mathbf{Z}_n^{GT}$ is a structured CoT that explains how reasoning over subtle acoustic and linguistic cues leads to the ambiguous emotion distribution, serving as an interpretable bridge between multimodal evidence and final predictions.

The objective is to learn a LALM $f_\theta$ that maps the input to a joint output space consisting of a predicted reasoning trajectory $\hat{\mathbf{Z}}_n$ and a predicted emotion distribution $\hat{\mathbf{p}}_n$: $(\hat{\mathbf{Z}}_n, \hat{\mathbf{p}}_n) = f_\theta(A_n, T_n)$. 
The model is optimized via two complementary objectives. It minimizes (i) the divergence between the predicted and ground-truth distributions while (ii) encouraging the generated reasoning steps to follow the ground truth reasoning trajectory $\mathbf{Z}_n$.

\vspace{-3pt}
\subsection{CoT Curation via Ambiguity-aware Prompting}
To enable the LALM to reason over ambiguous emotional cues, we first curate a structured CoT dataset that provides explicit reasoning supervision signals for ambiguity emotion. For each input $x_n = \{A_n, T_n\}$ and its corresponding ground-truth distribution $\mathbf{p}_n^{GT}$, we synthesize the reasoning trajectory $\mathbf{Z}_n^{GT}$ using a high-capacity, closed-source LALM (e.g., GPT-4o). 

The generation process follows a structured protocol to ensure the trajectory remains grounded and logically consistent as shown in Table~\ref{tab:cot_prompt}.
The generated trajectories are then automatically validated by checking whether the explanation is consistent with the given target distribution using the same closed-source model.

\begin{table}[t]
\centering
\vspace{-5pt}
\caption{Synthesis for ambiguity-aware CoT Curation.}
\label{tab:cot_prompt}
\fbox{
\begin{minipage}{0.95\linewidth}
\textbf{Input:}
$x_n = \{A_n, T_n\}$, $\mathbf{p}_n^{GT}$
\tightline
\textbf{Critical Rules for Reasoning Steps:}
\textbf{Keywords} Extract salient acoustic and linguistic cues.
\tightline
\textbf{Step 1: Text Analysis.} Analyze the text strictly for semantic meaning and context. Identify potential ambiguity.
\tightline
\textbf{Step 2: Audio Analysis.} Describe the prosody using professional terminology (Volume, Speed, Pitch, Tone). Highlight specific cues that support both the majority label AND the minority label (if any).
\tightline
\textbf{Step 3: Synthesis.} Synthesize the evidence to resolve the ambiguity. The reasoning must be strong enough that a reader would predict the Ground Truth labels just by reading your analysis.
\end{minipage}
}
\vspace{-10pt}
\end{table}

\vspace{-5pt}
\subsection{Ambiguity-aware Learning Objectives}
With the structured reasoning trajectory $\mathbf{Z}_n^{GT}$ and the perceptual distribution $\mathbf{p}_n^{GT}$, we explore the optimization of the LALM for ambiguity-aware emotion understanding. Specifically, we propose a Ambigutiy-Aware Objective designed to align the model outputs with human emotional uncertainty. A key advantage of our formulation is its ``plug-and-play" compatibility; it can be seamlessly integrated into various training frameworks, including SFT, DPO, and GRPO.

\vspace{-3pt}
\subsubsection{Ambigutiy-Aware Objective}
The goal is to align the model predicted emotion distribution $\hat{\mathbf{p}}_n$ with the human perceptual distribution $\mathbf{p}_n^{GT}$. We employ the Kullback-Leibler (KL) Divergence as the alignment criterion:
\begin{equation}
\mathcal{L}_{dist} = D_{KL}(\hat{\mathbf{p}}_n \parallel \mathbf{p}_n^{GT}) 
= \sum_{i=1}^{C} p_i^{GT} \log \frac{p_i^{GT}}{\hat{p}_i},
\end{equation}

Rather than relying on standard textual generation to obtain $\hat{\mathbf{p}}_n$, we read out token-level logits for emotion category names at the final step and applying a softmax over these logits to capture a more nuanced distribution of emotional states, following~\cite{halim2025token}. This logit-based formulation lets the model express graded uncertainty over emotion classes, mirroring the multi-annotator soft labels.

\subsubsection{Plug-and-Play Learning Paradigms}
Our distribution-aware objective is framework-agnostic, allowing for seamless integration into different training strategies. 

\vspace{5pt}
\noindent\textbf{SFT. }
For learning using SFT, the model is trained to jointly maximize the likelihood of the reasoning tokens and minimize the distributional divergence. The total loss $\mathcal{L}_{SFT}$ is a weighted combination:
\begin{equation}
\mathcal{L}_{SFT} = \mathcal{L}_{CE}(\hat{\mathbf{Z}}_n, \mathbf{Z}_n^{GT}) + \lambda \mathcal{L}_{dist}(\hat{\mathbf{p}}_n, \mathbf{p}_n^{GT}),
\end{equation}
where $\mathcal{L}_{CE}$ is the standard cross-entropy loss for the autoregressive generation of the CoT trajectory. This objective jointly constrains the reasoning trajectory and the distribution.

\vspace{3pt}
\noindent\textbf{DPO. }
For DPO, we extend the standard framework with our ambiguity-aware objective. Instead of relying on static, offline preference pairs, we use an on-policy scheme where the current policy generates \emph{dynamic} CoT rollouts $\hat{\mathbf{Z}}_n$ and their emotion distributions $\hat{\mathbf{p}}_n$. Rollouts whose predicted distribution deviates from the ground-truth distribution are treated as negative samples, while the curated ground truth CoT serves as the positive sample. This is measured by Jensen-Shannon (JS) Divergence, $JS(\hat{\mathbf{p}}_n, \mathbf{p}_n^{GT})$, which provides a bounded symmetric metric more stable for identifying ``poor" reasoning paths. The DPO loss then encourages preference for trajectories that both follow the gold reasoning and better match the human perceptual distribution, combined with the KL-based distribution loss and CoT cross-entropy as: 
\begin{equation}
\scriptsize
\mathcal{L}_{\text{DPO-total}} = \mathcal{L}_{\text{DPO}} + \lambda_1 \mathcal{L}_{dist}(\hat{\mathbf{p}}_n, \mathbf{p}_n^{GT}) + \lambda_2 \mathcal{L}_{CE}(\hat{\mathbf{Z}}_n, \mathbf{Z}_n^{GT}),
\end{equation}
\begin{equation}
\scriptsize
\mathcal{L}_{\text{DPO}}
=
- \mathbb{E}
\Big[
\log \sigma \Big(
\beta \big(
\log \tfrac{\pi_\theta(y_{pos}|x_n)}{\pi_{\text{ref}}(y_{pos}|x_n)}
-
\log \tfrac{\pi_\theta(y_{neg}|x_n)}{\pi_{\text{ref}}(y_{neg}|x_n)}
\big)
\Big)
\Big],
\end{equation}
where $\pi_\theta$ denotes the current LALM, and $\pi_{\text{ref}}$ denotes the original LALM without any update.

\vspace{3pt}
\noindent\textbf{GRPO. }
We additionally study reinforcement learning based reasoning optimization using GRPO, and similarly augment the reward with ambiguity-awareness.

At each update of the old policy $\pi_{\theta_{\text{old}}}$, we sample $K$ reasoning trajectories $\{\hat{Z}_{n,k}\}_{k=1}^{K} \sim \pi_{\theta_{\text{old}}}(\cdot|x_n)$ with corresponding emotion distributions $\{\hat{\mathbf{p}}_{n,k}\}_{k=1}^{K}$. Each trajectory is assigned a reward $r_{n,k}$:
\begin{equation}
    r_{n,k} =
    -\mathcal{L}_{dist}(\mathbf{p}_n^{GT}, \hat{\mathbf{p}}_{n,k})
    + \lambda_3 r_{\text{format}}(\hat{Z}_{n,k}),
\end{equation}
where the first term rewards accurate matching of the human perceptual distribution and the second term enforces adherence to the prescribed CoT format (defined similar as~\cite{diao2025soundmind}). We then normalize rewards into advantages to stabilize training: $\hat{A}_{n,k} = \frac{r_{n,k} - \mu_n}{\sigma_n}$, where $\mu_n$ and $\sigma_n$ denote the mean and standard deviation of advantages across the $K$ sampled trajectories for utterance $n$. 

The policy is then optimized using a GRPO-style objective:
\vspace{-5pt}
{ \footnotesize
\begin{align}
\mathcal{J}_{\text{GRPO}} & =
\mathbb{E}
\Bigg[
\frac{1}{K}
\sum_{k=1}^{K}
\frac{1}{|Z_{n,k}|}
\sum_{t}
\min\Big(
\rho_{n,k,t}(\theta)\hat{A}_{n,k}, \\ \nonumber
& \quad \text{clip}(\rho_{n,k,t}(\theta),1-\epsilon,1+\epsilon)\hat{A}_{n,k}
\Big)
- \beta \mathrm{KL}(\pi_\theta || \pi_{\text{ref}})
\Bigg].
\end{align}
}
where $\rho_{n,k,t}(\theta)$ denotes the token-level policy probability ratio. The objective increases the likelihood of high-quality reasoning trajectories while suppressing low-quality ones.

As GRPO typically relies on sampled trajectories without direct reference to ground-truth reasoning paths, we further enhance advantage estimation by incorporating the ground-truth trajectory $\mathbf{Z}_n^{GT}$ as an additional reference sample when advantages, referred to as GRPO$_z$. This ensures that the ground-truth reasoning trajectory consistently receives the highest reward, thereby biasing optimization toward more faithful reasoning paths under ambiguity.
\vspace{-1em}

\vspace{-3pt}
\section{Experimental Setup} 
\textbf{Datasets. }
We evaluate our training paradigm on IEMOCAP~\cite{busso2008iemocap} and CREMA-D~\cite{cao2014crema}. IEMOCAP contains four emotion categories: Anger, Happiness, Sadness, and Neutral, with “Excited” merged into Happiness, and is evaluated using strict 5-fold leave-one-session-out cross-validation. CREMA-D~\cite{cao2014crema} includes six categories, adding Disgust and Fear. 
We aggregate annotator votes (i.e., 3 annotators in IEMOCAP and 4-12 annotators in CREMA-D) 
and normalize them to construct soft ground-truth labels.

\noindent\textbf{Implementation Details. }
We use GPT-4o to construct structured CoT trajectories due to its strong capability in generating coherent reasoning chains that connect input observations with target outputs~\cite{zhifei2025audio}. 
Experiments with post-training are conducted using Qwen2-Audio-7B-Instruct\footnote{\url{https://huggingface.co/Qwen/Qwen2-Audio-7B-Instruct}} as the base LALM. 
For all the learning paradigms, we apply LoRA ($r = 8$, $\alpha_{\text{lora}} = 16$) to the attention and feed-forward modules. All trainable parameters are optimized using AdamW with optimized learning rates of 1e-4 (SFT), 5e-6 (DPO), and 2e-5 (GRPO). A 3\% linear warmup and cosine decay are applied throughout training. The corresponding loss weights are $\lambda = 1, \lambda_1 = 0.1$, $\lambda_2 = 0.1$, and $\lambda_3 = 0.1$  for regularization during DPO optimization. 
We compare against the 1) Base Model and 2) Audio-Reasoner~\cite{zhifei2025audio}.

\noindent\textbf{Evaluation Metrics. }
To assess how well the predicted emotion distribution matches the ground-truth distribution, we report Jensen–Shannon divergence (JS $\downarrow$) and the Bhattacharyya Coefficient (BC $\uparrow$)~\cite{hong2025aer}, which measure distributional discrepancy and probability mass overlap, 
 assessing if the model preserves uncertainty rather than overconfident predictions.

\begin{table}[t]
\centering
\setlength{\tabcolsep}{3pt}
\caption{Comparison for ambiguous emotion recognition performance. 
        The best results are in \textbf{bold}. GRPO$_z$ indicates ground-truth CoT trajectory injection in GRPO.}
\begin{tabular}{l l c c c c}
\toprule
Dataset & Methods & JS$\downarrow$ & BC$\uparrow$ & $R^2$$\uparrow$ & Brier$\downarrow$ \\
\midrule
\multirow{6}{*}{IEMOCAP}
& Base model & 0.40 & 0.64 & 0.51 & 0.15 \\
& Audio-Reasoner & 0.36 & 0.67 & 0.52 & 0.15 \\
\cmidrule(l){2-6}
& SFT & 0.29 & 0.73 & 0.60 & 0.12 \\
& DPO & 0.23 & 0.79 & 0.65 & 0.09 \\
& GRPO & 0.36 & 0.67 & 0.55 & 0.13 \\
& GRPO$_z$ & \textbf{0.20} & \textbf{0.82} & \textbf{0.67} & \textbf{0.07} \\

\midrule
\multirow{6}{*}{CREMA-D}
& Base model & 0.25 & 0.78 & 0.54 & 0.05 \\
& Audio-Reasoner & 0.37 & 0.69 & 0.51 & 0.09 \\
\cmidrule(l){2-6}
& SFT & 0.21 & 0.82 & 0.57 & 0.04 \\
& DPO & \textbf{0.17} & \textbf{0.86} & 0.67 & \textbf{0.03} \\
& GRPO & 0.25 & 0.77 & 0.46 & 0.04 \\
& GRPO$_z$ & 0.20 & 0.82 & \textbf{0.68} & 0.04 \\
\bottomrule
\end{tabular}
\label{tab:main}\vspace{-10pt}
\end{table}

\vspace{-3pt}
\section{Results}

\subsection{Performance of Post-Training Strategies}

We first evaluate the effectiveness of the proposed ambiguity-aware training paradigm across different post-training strategies in Table~\ref{tab:main}. 
Across both datasets, applying the proposed objectives consistently improves performance over the Base model under SFT, DPO, and GRPO$_Z$. 

Among them, SFT consistently performs worse than both GRPO$_Z$ and DPO, suggesting that ambiguity-aware emotion modelling benefits from learning over multiple reasoning trajectories rather than relying on a single supervised reasoning path. Notably, GRPO$_Z$ achieves the strongest performances on IEMOCAP across all metrics, whereas DPO performs best on CREMA-D. This difference suggests that the effectiveness of post-training strategies depends on the structural complexity of the emotion distribution space. As the number of emotion classes increases in CREMA-D, the predicted distribution becomes more sensitive to subtle variations in intermediate reasoning steps. In such higher-dimensional ambiguity settings, the outcome-level reward signal used in GRPO$_Z$ relies largely on the divergence between the predicted and target emotion distributions at the final decision stage 
, which may become less precise for guiding fine-grained reasoning. In contrast, preference-based methods such as DPO provide denser supervision by directly contrasting positive and negative reasoning chains at the token level, leading to stronger intermediate trajectory supervision toward the target distribution.



\vspace{-3pt}
\subsection{Impact of KL divergence}

\begin{figure}[t]
    \centering
    \includegraphics[width=\columnwidth]{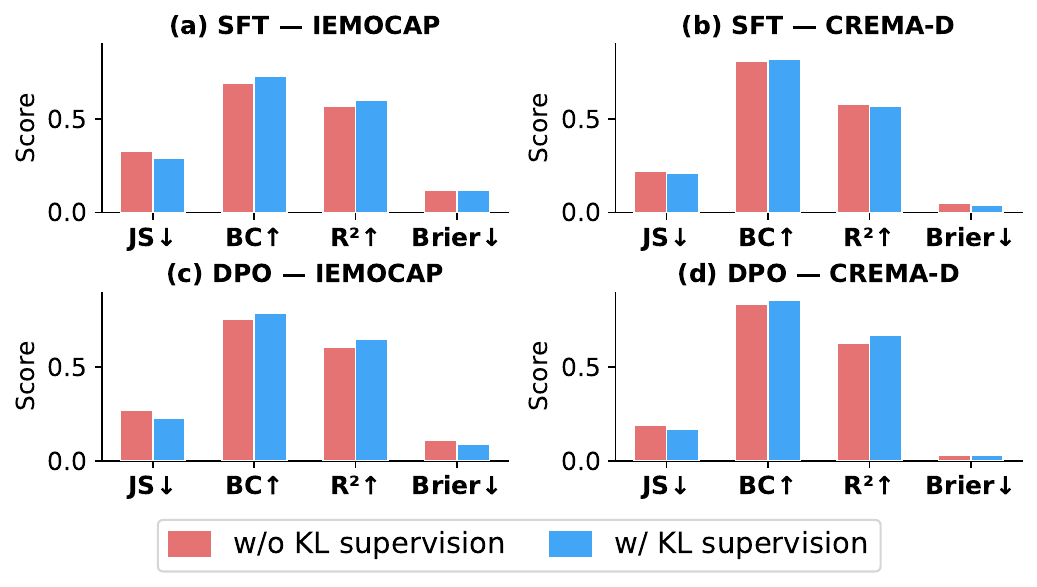}
    \vspace{-10pt}
    \caption{Comparison between CE-only (w/o KL supervision) and proposed additional KL supervision. (a) SFT on IEMOCAP, (b) SFT on CREMA-D, (c) DPO on IEMOCAP, and (d) DPO on CREMA-D.}
    \label{fig:CE_KL}\vspace{-10pt}
\end{figure}


We analyze the impact of KL-based distribution supervision by comparing variants trained with cross-entropy (CE) only against those incorporating KL divergence for emotion distribution supervision in Figure~\ref{fig:CE_KL}.


Across both datasets, incorporating KL supervision consistently improves distribution-level metrics compared to CE-only training. On IEMOCAP, adding KL regularization reduces JS divergence and increases BC for both SFT and DPO variants relative to their CE-only counterparts. Similar improvements are observed on CREMA-D. These consistent gains indicate that while CE-based training encourages the model to reproduce the curated reasoning trajectory and final prediction tokens, it performs weak supervision to explicitly constrain the predicted probability mass to match the soft-label emotion distribution. In contrast, KL divergence directly penalizes distribution mismatches, leading to improved consistency with ambiguity. 

\vspace{-3pt}
\subsection{Impact of ambiguity-aware CoT supervision}
\begin{table}[t]
\centering
\setlength{\tabcolsep}{5pt}
\caption{Ablation of SFT training objectives. Models are trained on CREMA-D and evaluated under in-domain (CREMA-D) and cross-domain (IEMOCAP) settings.}
\vspace{-5pt}
\label{tab:sft_ablation}
\begin{tabular}{lcc|cccc}
\toprule
Setting & CoT & KL & JS$\downarrow$ & BC$\uparrow$ & $R^2\uparrow$ & Brier$\downarrow$ \\
\midrule
In-domain & \cmark & \cmark & 0.21 & 0.82 & 0.57 & 0.04 \\
In-domain & \xmark & \cmark & 0.22 & 0.81 & 0.57 & 0.04 \\
\midrule
Cross-domain & \cmark & \cmark & 0.38 & 0.65 & 0.50 & 0.15 \\
Cross-domain & \xmark & \cmark & 0.52 & 0.52 & 0.31 & 0.22 \\
\bottomrule
\end{tabular}
\vspace{-2pt}
\end{table}

We analyze the impact of ambiguity-aware CoT supervision in Table~\ref{tab:sft_ablation}. Comparing SFT trained with and without CoT supervision reveals different behaviors under in-domain and cross-domain evaluation. When training and testing on CREMA-D, incorporating CoT supervision yields marginal performance gain comparing to KL-only training. However, when training on CREMA-D and testing on IEMOCAP, models trained with CoT supervision significantly outperform those trained without. This observation suggests that optimizing only the final emotion distribution through KL regularization without CoT can easily overfit to dataset-specific distribution patterns. In contrast, ambiguity-aware CoT supervision encourages the model to reason over multimodal emotional cues, leading to stronger generalization capability.
\vspace{-4pt}
\section{Conclusions} 


This work presents the first systematic study of ambiguity-aware reasoning in large audio–language models for ambiguous speech emotion recognition. We reformulate ambiguous emotion understanding as a distributional reasoning problem and propose a plug-and-play framework that combines an ambiguity-aware objective with structured chain-of-thought supervision. Different post-training strategies, such as SFT, DPO, and GRPO, using our training paradigm consistently improve performance on both IEMOCAP and CREMA-D. 

\section{Generative AI Use Disclosure}
Generative AI tools were used solely for minor language editing and polishing to improve the clarity and readability of the manuscript. These tools were not used to generate scientific content, analyze experimental results, summarize related work, develop methodologies, or propose research ideas. All conceptual contributions, experimental design, analysis, and conclusions presented in this paper were developed and verified by the authors. The authors take full responsibility for the content of this manuscript.

\bibliographystyle{IEEEtran}
\bibliography{mybib}

\end{document}